\documentclass[e-only,10pt,reqno]{ofj}

\usepackage{tensorNotation}
\usepackage{finiteVolumeNotation}

\usepackage{setspace}
\usepackage{lineno}
\usepackage{color}
\usepackage[symbol]{footmisc}
\usepackage[dvipsnames,svgnames,x11names]{xcolor}
\usepackage{orcidlink}

\let\paragraph\undefined

\copyrightinfo{2021}{OpenFOAM\textsuperscript{\textregistered} Journal}
\usepackage{subfigure}
\usepackage{caption}
\begin{document}

%-----------------------------------------------------------------------
%      Manuscript title and author description
%-----------------------------------------------------------------------

%    \title[short text for running head]{full title}
\title[Running Python codes in OpenFOAM using pybind11]{A general approach for running Python codes in OpenFOAM using an embedded pybind11 Python interpreter}

%    Author information
%    Note: authors should not be defined for the review process as the review is double-blind
%    Author names should be given as initials for forenames followed by the surname: see the examples below


%    \author[short version for running head]{name for top of paper}

%    Corresponding author information
%    Indicate the corresponding author with an asteriks
%    You can optionally add an Orcid link
\author{Simon Rodriguez$^{1,*}$\orcidlink{0000-0001-6924-2428}}
\address{School of Mechanical and Materials Engineering, University College Dublin, Ireland$^1$.}
\email{simon.rodriguezluzardo@ucdconnect.ie}

%    Author two information
\author{Philip Cardiff$^1$}
%\address{$2$Address2} % not needed if the same as author 1
% \email{ simon.rodriguezluzardo@ucdconnect.ie}

%   Author three information
%\author{P. Murphy$^2$}
%\address{$^2$Address2}
%\email{Emailaddress2}

%-----------------------------------------------------------------------
%      Manuscript abstract
%-----------------------------------------------------------------------

%    Abstract is required
\begin{abstract}
As the overlap between traditional computational mechanics and machine learning grows, there is an increasing demand for straight-forward approaches to interface Python-based procedures with C++-based OpenFOAM.
This article introduces one such general methodology, allowing the execution of Python code directly within an OpenFOAM solver without the need for Python code translation.
The proposed approach is based on the lightweight library pybind11, where OpenFOAM data is transferred to an embedded Python interpreter for manipulation, and results are returned as needed.
Following a review of related approaches, the article describes the approach, with a particular focus on data transfer between Python and OpenFOAM, executing Python scripts and functions, and practical details about the implementation in OpenFOAM.
Three complementary test cases are presented to highlight the functionality and demonstrate the effect of different data transfer approaches: a Python-based velocity profile boundary condition; a Python-based solver for prototyping; and a machine learning mechanical constitutive law class for solids4foam which performs field calculations.
\end{abstract}

% \date{\today}

\dedicatory{}

\maketitle
%   Line numbering should be enable for review and disabled for final version
% \linenumbers

%-----------------------------------------------------------------------
%      Start introduction here and continue with additional sections
%-----------------------------------------------------------------------

\section{Introduction}
%--------------------------------------------------------------------------------------%
%--------------------------------------------------------------------------------------%

Appearing in 1985, C++ was motivated by the limitations of the structured programming paradigm: code reusability, code sharing, and code extensibility \cite{sharma2014object}.
Built on code abstraction and pioneering the use of object-oriented programming (OOP), C++ quickly became the number one choice for the creation of large, complex, and extendable software.
In fact, before C++, OOP was mostly unknown in industry as its techniques were believed to be excessively expensive for real-world applications and too complex for non-expert programmers \cite{stroustrup2014programming}.
OpenFOAM was one of the first large computational mechanics softwares to embrace C++ and OOP, tracing its origin to its predecessor `FOAM' with its roots in the late 80's.
Using OOP techniques, the authors of FOAM represented computational entities using mathematical language, without sacrificing computational efficiency \cite{Weller1998}.

Over the subsequent three decades, there has been an explosion in the availability of computational resources, in particular via parallel computing, as well as C++ software to exploit it.
However, as acknowledged by Bjarne Stroustrup when creating C++, code elegance was not its primary goal: \emph{``Even I knew how to design a
prettier language than C++''}  \cite{stroustrup2014programming}.
The aim for C++ was utility, not abstract beauty. 
In contrast, Python, publicly released in 1991, aimed to achieve a higher level of abstraction which emphasised readability and usability, while achieving it with simpler syntax and fewer rules.
As a result, when compared to lower-level languages such as C++, Python's learning curve is less steep and requires considerably less effort for prototyping code and performing auxiliary tasks such as handling files;
however, these benefits are typically achieved at a higher computational cost \cite{mckinney_python_2018}.

Over time, Python has become one of the most widely used programming languages, with the `Popularity of Programming Languages Index' \cite{noauthor_pypl_nodate} placing it
as the most popular programming language in the world.
This momentum has resonated with the OpenFOAM community, which has tried to merge the advantages of Python with OpenFOAM, as demonstrated by projects such as:
\begin{itemize}
%\tightlist
\item  PyFoam: A widely used library for manipulating OpenFOAM cases and controlling OpenFOAM runs \cite{noauthor_pyfoam_nodate};
\item PythonFlu: a Python-based OpenFOAM wrapper which aims to make the calculation environment more interactive and automated \cite{noauthor_pythonflu_nodate}; however, in recent years this project has become inactive;
\item fluidfoam: Python classes for plotting OpenFOAM data \cite{cyrillebonamy_fluidfoam_nodate};
\item  Owls: Python tools for data analysis and plotting of OpenFOAM cases \cite{noauthor_owls_nodate}.
\end{itemize}
In general, these projects have used Python to improve the user experience or develop pre/post processing OpenFOAM utilities.
On the other hand, swak4Foam \cite{noauthor_swak4foam_nodate}, which is a collection of parsers for expressions on OpenFOAM-types, was designed with the goal of minimising the use of C++ in OpenFOAM.
With this premise, one of its functionalities offers a function object that can execute Python code using an integrated Python interpreter.
In addition, Maulik et al. \cite{maulik2021pythonfoam} and Weiner et al. \cite{weiner_datadriven_2019, andre_weiner_running_nodate} have aimed to provide capabilities to use data analysis tools from Python within OpenFOAM;
the former constructed OpenFOAM applications that have bindings to data libraries in Python and the latter transfers data and neural network models from Python to OpenFOAM via the PyTorch library.
More recently, Anderluh and Jasak \cite{anderluh_jasak_like_2021} have proposed to re-write the core OpenFOAM functionality entirely in Python.

Inspired by swak4Foam, the current article proposes a general approach for integrating an embedded Python interpreter in OpenFOAM via the pybind11 library \cite{pybind11}, with a particular emphasis on its use with machine learning techniques.
In particular, the approach is motivated by two opportunities:
\begin{enumerate}
\item	The ability to rapidly prototype numerical algorithms using Python, while taking advantage of the OpenFOAM framework;

\item The capacity to exploit the vast ecosystem of scientific programming tools that have been developed for Python, particularly those in the area of machine learning and data science.
\end{enumerate}

Related to the second point, the recent machine learning revolution has lead to the nascent area of hybrid machine-learning-computational-mechanics procedures, including
numerous research avenues: machine learning-acceleration of solvers \cite{kochkov2021machine, obiols-sales_cfdnet_2020};  machine learning-constitutive laws in solid mechanics \cite{pandya_computational_2017, abueidda_deep_2021, wang_general_2020}; machine learning-turbulence models in fluid dynamics \cite{ling2016reynolds}; and targeting both acceleration and accuracy gains \cite{pop_deep_2020}.
In these approaches, the first challenge involves the development of a machine learning procedure - typically based on a neural network - capable of solving the required task; this is commonly done with Python, or similar languages such as R, Matlab or Julia.
The next immediate challenge consists of coupling the machine learning model with a computational mechanics solver, whether OpenFOAM or another; currently, there is no standard way of doing this in OpenFOAM. Possible approaches include:
\begin{enumerate}
	\item Using the TensorFlow C++ API where OpenFOAM is linked at compile time to TensorFlow \cite{maulik2021deploying};
	
	\item Using the Python C API combined with the NumPy package C API \cite{maulik2021pythonfoam};
	
	\item Translating the Python-based neural network to C++ code via specialised conversion tools, such as frugally-deep \cite{tobias_hermann_frugally-deep_nodate}.
\end{enumerate}

These solutions all require the Python code and/or models to be translated or converted to allow their use with C++ in OpenFOAM.
The primary conceptual disadvantage with this is that it conflicts with the main motivation of many programmers who want to use Python: coding simplicity;
for example, this is particularly evident when attempting to compile and link the TensorFlow C++ API with OpenFOAM, where both must be built from source with consistent settings.
These challenges can limit the adoption of Python as a development tool for applications beyond traditional machine learning for which Python might be well-suited;
individual components of OpenFOAM solvers that are particularly suitable are field calculations, boundary conditions, function objects, and pre/post processing utilities.

Motivated by this, the current article presents a general approach to execute Python code from within OpenFOAM without the need to convert the Python code to C++.
This approach uses an embedded Python interpreter based on the lightweight library pybind11 \cite{pybind11}.
The approach, which is similar to the approaches adopted in swak4Foam \cite{noauthor_swak4foam_nodate} and by Maulik et al. \cite{maulik2021pythonfoam}, enables the use of Python classes/functions/libraries via their ``pure'' Python interfaces, not lower level ones.

The remainder of this article is structured as follows:
Section 2 presents an introduction to the required pybind11 support including a general description on pybind11 and how to include it in OpenFOAM;
this is followed by a description of the interaction between OpenFOAM and Python.
In Section 3, three different test cases are used to illustrate how the proposed approach can be used in OpenFOAM.
The first case demonstrates how to implement a general velocity profile boundary condition via Python.
The second case shows how the Python interpreter can be used to quickly test ``proof of concept'' Python numerical implementations via an OpenFOAM solver.
The final case gives an example of how to perform field calculations via analytical calculations in Python as well as via TensorFlow/Keras machine learning models in Python;
in this case, an elastic mechanical constitutive law for a solid mechanics simulation is chosen for demonstration, but the approach is equally applicable to any other field calculation.
Section 4 summarises the main findings and learning points of the article, and briefly discusses future directions.
%and alternative related implementations.

%--------------------------------------------------------------------------------------%
%--------------------------------------------------------------------------------------%

\section{Approach and implementation} \label{sec:approach}
This section gives a brief description of pybind11, followed by steps on how to use it with OpenFOAM.
After explaining the high level interaction between OpenFOAM and Python via pybind11, the core technical steps are outlined:
\begin{itemize}
	\item Initialising the Python interpreter;
	\item Transferring data between OpenFOAM and the Python interpreter;
	\item Executing Python code from OpenFOAM.
\end{itemize}

%--------------------------------------------------------------------------------------%
\subsection{The pybind11 library} {\label{sec:pybind_library}}
%--------------------------------------------------------------------------------------%
pybind11 is a lightweight header-only library that is primarily used to allow C++ code to be called from Python \cite{pybind11}; however, as used in this article, pybind11 also offers the possibility of embedding a Python interpreter in a C++ application.

%--------------------------------------------------------------------------------------%
\textbf{Installing pybind11 with Conda}
%--------------------------------------------------------------------------------------% 
Conda is an open source package and environment management system, which can be used for managing Python environments.
Using Conda, pybind11 can be installed in a *nix terminal with:
\begin{lstlisting}[]
$> conda install -c conda-forge pybind11
\end{lstlisting}

%--------------------------------------------------------------------------------------%
\textbf{Including pybind11 in OpenFOAM} 
%--------------------------------------------------------------------------------------%
Once pybind11 is installed on the system, it can be used in an OpenFOAM application or library by appropriate modification of the OpenFOAM \texttt{Make/options} file:

\begin{itemize}

\item  Declare two variables, \texttt{PYTHON\_INC} and \texttt{PYTHON\_LIB\_DIR}, at the top of the \texttt{options} file to store the location of the required pybind11 header files and the Python libraries:

\begin{lstlisting}[]
PYTHON_INC_DIR := $(shell python3 -m pybind11 --includes)
PYTHON_LIB_DIR := $(shell python3 -c "from distutils import sysconfig; \
    import os.path as op; v = sysconfig.get_config_vars(); \
    fpaths = [op.join(v[pv], v[`LDLIBRARY']) for pv in (`LIBDIR', `LIBPL')]; \
    print(list(filter(op.exists, fpaths))[0])" | xargs dirname)
\end{lstlisting}

where the shell commands here use the \texttt{python3} command to lookup the location of the include and library directories. 

\item  Include \texttt{PYTHON\_INC\_DIR} in the \texttt{EXE\_INC} field, for example:

\begin{lstlisting}[]
EXE_INC = \
    $(PYTHON_INC_DIR)
\end{lstlisting}

\item Include \texttt{PYTHON\_LIB\_DIR} and the Python C dynamic library in the \texttt{EXE\_LIBS} field (or in the \texttt{LIB\_LIBS} field if compiling a library), for example:

\begin{lstlisting}[]
EXE_LIBS = \
    -L$(PYTHON_LIB_DIR) \
    -lpython3.8
\end{lstlisting}
\end{itemize}
where in this case, Python version 3.8 is used, but the programmer should change this to whichever version they are using.

%--------------------------------------------------------------------------------------%
\textbf{Including the required header files} 
%--------------------------------------------------------------------------------------%
The header files that will be required in the OpenFOAM/Python communication depend on the specific interaction scheme used by the programmer;
however, in general the following header files should be included in the OpenFOAM source code:
\begin{lstlisting}[language=C++]
#include <pybind11/embed.h>
#include <pybind11/eval.h>
#include <pybind11/stl.h>
#include <pybind11/numpy.h>
\end{lstlisting}

These four header files provide the following functionality:
\begin{itemize}
	\item \texttt{embed.h}: Enables the functionality required to embed the Python interpreter in a C++ program;
	\item \texttt{eval.h}: Allows Python code to be executed/evaluated from C++ in the embedded Python interpreter;
	\item \texttt{stl.h}: Tools for converting data from C++ standard template library containers, such as \texttt{std::vector}, to Python arrays and vice versa;
	\item \texttt{numpy.h}: NumPy is the workhorse of scientific computation in Python, and although it is not strictly required, it is often convenient and efficient to perform Python computations via NumPy. This header allows to use the \texttt{py::array} data type, which is a C++ wrapper for NumPy arrays. When a \texttt{py::array} is sent to the Python interpreter, the wrapper is removed and the data is exposed as a NumPy array.
\end{itemize}

Assuming pybind11 was installed correctly, these header files should all be available within the \texttt{PYTHON\-\_INC\-\_DIR} directly defined in the \texttt{Make/options} file.

%--------------------------------------------------------------------------------------%
\textbf{Note on \texttt{setRootCase.H}}
%--------------------------------------------------------------------------------------%
OpenFOAM applications typically include a header file named \texttt{setRootCase.H} which is used to read and check command-line arguments;
however, the default arguments which are used when constructing the \texttt{Foam::argList} object within \texttt{setRootCase.H} can cause issues with certain Python modules, such as TensorFlow, as noted in online forums \cite{noauthor_question_nodate}.
This can be solved by modifying the default \texttt{setRootCase.H} file in the following way:
\begin{lstlisting}[language=C++]
// Foam::argList args(argc, argv); // previous line
Foam::argList args(argc, argv, true, true, false);
if (!args.checkRootCase())
{
    Foam::FatalError.exit();
}
\end{lstlisting}

%--------------------------------------------------------------------------------------%
\subsection{General workflow for using pybind11 with OpenFOAM} {\label{776699}}
%--------------------------------------------------------------------------------------%
The central idea revolves around using pybind11 to create a Python interpreter that exists alongside the OpenFOAM application. Then, pybind11 allows data to be transferred to and from the Python interpreter, as well as the ability to execute Python functions and scripts from within OpenFOAM.

\begin{figure}[htb]
	\begin{center}
	\includegraphics[width=0.90\columnwidth]{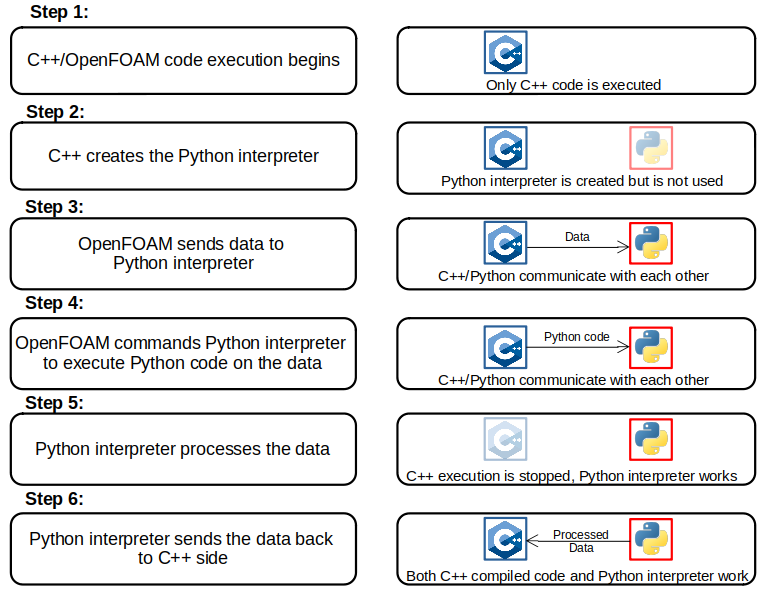}
	\caption{{Overview of the interaction between OpenFOAM and Python, via pybind11. {\label{InteractionOpenFOAMPython}}} 
	 }	
	\end{center}
\end{figure}

The entire process is conceptually depicted in Figure {\ref{InteractionOpenFOAMPython}}.
OpenFOAM execution begins as usual, for example with common tasks involving the creation and initialisation of fields and other auxiliary data structures (Step 1).
At a convenient point, a Python interpreter is created and held in the background (Step 2).
The OpenFOAM execution continues and, when required, it sends data to the Python interpreter (Step 3).
An OpenFOAM command is then called to execute a piece of Python code (Step 4) which invokes the Python interpreter (Step 5); while the Python interpreter is executing, the OpenFOAM code waits.
Finally, the data is sent back to OpenFOAM, where it can be used in the rest of OpenFOAM workflow (Step 6).
Details of the different steps shown in Figure {\ref{InteractionOpenFOAMPython}} are given in the subsequent sub-sections.

\subsection{Initialising the Python interpreter} {\label{966233}}
To improve readability, it is assumed that all subsequent OpenFOAM code has invoked the C++ namespace \texttt{pybind11} as shown:
\begin{lstlisting}[language=C++]
namespace py = pybind11;
\end{lstlisting}

To initialise, embed and expose the Python interpreter within OpenFOAM, the following line is used:
\begin{lstlisting}[language=C++]
py::initialize_interpreter();
\end{lstlisting}
This can be considered a higher level version of the \texttt{Py\_Initialise()} method from the Python C API and its location in the code must ensure that it is run only once. After the interpreter is created, it can be interacted with at any time.
The line can be placed anywhere convenient within the OpenFOAM code, for example, in the \texttt{createFields.H} header file of a solver, in the constructor of a class, or whenever it is needed.

%--------------------------------------------------------------------------------------%
\subsection{Transferring data between OpenFOAM and Python} \label{TransferOFPython}
%--------------------------------------------------------------------------------------%

To facilitate data transfer between OpenFOAM and Python, a Python scope object must be created and stored in OpenFOAM:
\begin{lstlisting}[language=C++]
py::object scope = py::module_::import("__main__").attr("__dict__");
\end{lstlisting}

A dissection this line requires analysing both \texttt{\_\_main\_\_ }and \texttt{\_\_dict\_\_} in a Python context.
According to the official Python documentation \cite{noauthor___main___nodate}, \texttt{\_\_main\_\_} is the name of the environment where top-level code is run.
It is top-level because it imports all other modules that the program needs.
\texttt{\_\_dict\_\_}, on the other hand, is a dictionary or other mapping object used to store an object's (writable) attributes \cite{noauthor_built-types_nodate}.
Therefore, this line imports the dictionary that contains all the variables defined in the Python interpreter and assigns it to the object \texttt{scope} on the C++ side;
this C++ scope object can be used to declare variables in the Python interpreter from C++ or to retrieve variables from Python to the C++ side.
 
%--------------------------------------------------------------------------------------%
\textbf{Passing by copy} \label{passingByCopy}
%--------------------------------------------------------------------------------------%
Using this \texttt{scope} object, it is possible to copy the variables from OpenFOAM to Python;
for example, the OpenFOAM scalar variable \texttt{a} can be copied to a new variable in the Python scope, \texttt{x}, as follows:
\begin{lstlisting}[language=C++]
const scalar a = 2;
scope["x"] = a;
\end{lstlisting}
Similarly, the \texttt{scope} object can be used to copy variables from the Python scope to OpenFOAM:
\begin{lstlisting}[language=C++]
const scalar b = scope["y"];
\end{lstlisting}
where it is assumed here that a scalar variable ``y'' exists in the Python scope before this line is executed.

%--------------------------------------------------------------------------------------%
%\paragraph{Casting Python to C++ data types and vice versa} 
%--------------------------------------------------------------------------------------%
The examples above show how primitive data types (bool, integer, float, double, etc) can be transferred between OpenFOAM and Python;
this approach can be easily applied to OpenFOAM data containers, such as \texttt{List} and \texttt{Field}, by passing each element of these containers one-by-one to the Python interpreter to be processed. The following example code demonstrates how this approach could be used to perform cell-by-cell calculations using the pressure field \texttt{p}:

\begin{lstlisting}[language=C++]
const scalarField& pI = p.primitiveField();
scalarField result(pI.size(), 0.0);
forAll(p, cellI)
{
    // Transfer p for cellI to the Python interpreter
	scope["p"] = pI[cellI];
        
    // Perform some calculation in Python as described in Section 2.5}
    // ...

    // If applicable, retrieve data from Python and store it in an OpenFOAM container
    // For example:
    result[cellI] = scope["result"];
}
\end{lstlisting}

Although the approach above works, it may be prohibitively slow for larger fields; the reason for this is that compiler optimisations on the OpenFOAM side are inhibited, and, similarly, efficient NumPy array operators are not utilised on the Python side.
An alternative approach is to pass the entire field to the Python interpreter, process the entire field in Python, and transfer the result field back to OpenFOAM.
To do this, we will use pybind11 wrapper classes as described below.

As stated in the pybind11 documentation, \emph{``all major Python types are available as thin C++ wrapper classes}" \cite{noauthor_pythontypes_nodate}, including \texttt{object},  \texttt{bool}, \texttt{int}, \texttt{float}, \texttt{str}, \texttt{bytes}, \texttt{tuple}, \texttt{list} and others.
In general, to convert C++ types to Python, the method \texttt{py::cast()} is used \cite{pybind11}.
For example, if a \texttt{std::vector\textless{}double\textgreater{}} named \texttt{inputC} has been declared and initialised in OpenFOAM/C++,
it is possible to create a C++ object \texttt{inputPy} of type \texttt{py::array} which can be directly transferred to the Python interpreter as a NumPy array.
The following example demonstrates this behaviour:
\begin{lstlisting}[language=C++]
std::vector<scalar> x{10, 20, 30};
py::array pyX = py::cast(x);
scope_["x"] = pyX;
\end{lstlisting}
where the \texttt{std::vector} is copied to a C++ \texttt{py::array} which can be directly transferred to the Python scope.
The same conversions are allowed when the arrays are multi-dimensional, for example, 2-D arrays;
the only difference in that case is that the \texttt{x} variable must be declared as a vector of vectors, i.e. \texttt{std::vector\textless{}std::vector\textless{}type\textgreater{}\textgreater{}}.
The reverse operation, from a Python NumPy array to a C++ \texttt{std::vector}, uses similar syntax:
\begin{lstlisting}[language=C++]
const py::array resultPy = scope["result"];
const std::vector<scalar> result = resultPy.cast<std::vector<double>>();
\end{lstlisting} 
If the NumPy array has two dimensions it would be cast to a \texttt{std::vector\textless{}std::vector\textless{}type\textgreater{}\textgreater{}.}

%--------------------------------------------------------------------------------------%
%\subsection{Processing OpenFOAM fields in the Python interpreter} {\label{794249}}
%\paragraph{Processing OpenFOAM fields in the Python interpreter} {\label{794249}}
%--------------------------------------------------------------------------------------%

\begin{figure}[htb]
     \centering
     \begin{subfigure}
         \centering
         \includegraphics[width=\textwidth]{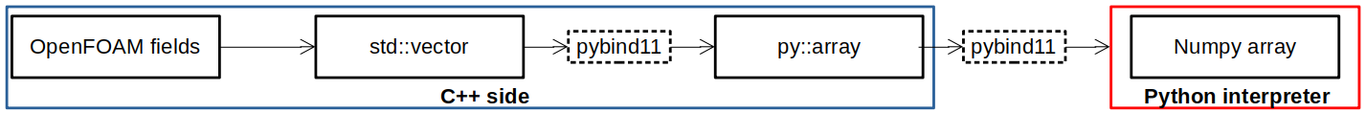}
         \caption{Forward conversion. Passing an OpenFOAM field to the Python interpreter implies extracting its data to create a C++ std::vector, creating a py::array from it and transferring it to the Python interpreter}
         \label{fig:ForwardConversion}
     \end{subfigure}
     \vfill
     \begin{subfigure}
         \centering
         \includegraphics[width=\textwidth]{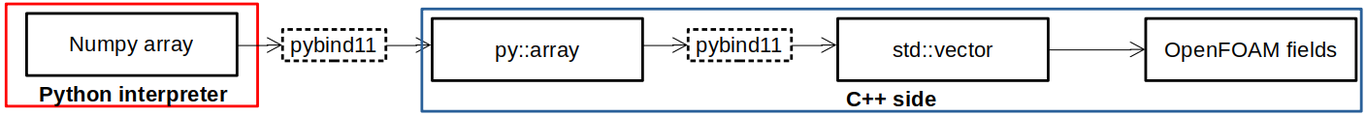}
         \caption{Reverse (backward) conversion. Passing an NumPy array from the Python interpreter implies transferring it to a py::array in the C++ side, which is later cast to a std::vector via pybind11 and finally converted back to OpenFOAM field}
         \label{fig:ReversConversion}
     \end{subfigure}
\end{figure}

In the examples above, \texttt{std::vector} data containers are used on the C++ side; however, as \texttt{Field} and \texttt{List} containers are the standard in OpenFOAM, we explain here how to convert between these container types.
Essentially, the \texttt{std::vector} containers are used as a intermediate step when transferring field data between OpenFOAM and Python, as illustrated schematically in Figure \ref{fig:ReversConversion}.

The implication of passing OpenFOAM field data on an element-by-element basis or as an entire field is examined in the third and final test case in Section \ref{sec:test_case_field_calculations}.

%--------------------------------------------------------------------------------------%
\textbf{Passing by reference} 
%--------------------------------------------------------------------------------------%
Using the approach described above, a copy of the data is created when it is transferred between C++ and Python; that is to say, when a C++ variable is transferred to the Python interpreter scope, the C++ data type is copied to a Python data type.
As an alternative, it is possible to exchange data between C++/OpenFOAM and Python by \emph{reference} rather than as a copy.
The approach is based on the built-in NumPy support for \texttt{ctypes}, which \emph{``provides C compatible data types''} in Python \cite{noauthor_ctypes_nodate}.
The idea here is to introduce minor changes in the original Python code to allow Python to interact directly with the underlying data in the OpenFOAM C++ code, thus removing the need for data copying steps.
Every OpenFOAM field stores its data in the form of a C array, and a pointer to the first element of this array can be retrieved using the \texttt{data} (or \texttt{cdata}) functions provided by the underlying \texttt{UList} base class.
Using pybind11, this pointer address is converted to an unsigned integer and sent to the Python interpreter along with the size/length of the array.
From the Python side, this integer address and size are used to create a NumPy array, which provides a reference to the C array data, without copying data.
This can be achieved in Python as:
\begin{lstlisting}[language=C++]
strain_pointer = ctypes.cast(strain_address, ctypes.POINTER(ctypes.c_double))
strain = np.ctypeslib.as_array(strain_pointer, shape = (SIZE, 6))
\end{lstlisting} 
where \texttt{strain\_address} is the integer address from C++/OpenFOAM and \texttt{SIZE} is the size/length of the array.
Since the arrays created in the Python side are directly manipulating the underlying data of the OpenFOAM fields, all the modifications done in the Python side are immediately propagated to the OpenFOAM side, eliminating the need for copying the data.
Clearly, care must be taken when using this approach as memory errors may occur if the address changes on the C++ side without informing Python.

%--------------------------------------------------------------------------------------%
\subsection{Executing Python code from OpenFOAM} \label{sec:execute_python_code}
%--------------------------------------------------------------------------------------%
The command \texttt{py::eval\_file(string, scope)} where \texttt{string} is the name of a Python file/script, loads all the entities defined in the Python file, which can then be consumed using the pybind11 \texttt{py::exec} command. 
For example, assume the following trivial Python function exists in the loaded Python script:
\begin{lstlisting}[language=Python]
def double_value(number):
    return 2.0*number
\end{lstlisting} 
This function receives a number as an argument and returns the number $\times$ 2. The following C++ code demonstrates how to use this Python function:
\begin{lstlisting}[language=C++]
// Define scalar `a' in OpenFOAM
const scalar a = 2.0;
// Transfer OpenFOAM variable `a' to a variable `x' in the Python interpreter
scope["x"] = a;
// Execute the Python function `double_value'
py::exec("y = double_value(x)", scope);
// Transfer the value of the Python variable `x' to the OpenFOAM scalar `b'
const scalar b = scope["y"];
\end{lstlisting} 

\begin{itemize}
%\tightlist
	\item Declares a \texttt{scalar} variable \texttt{a} and initialises its value to 2.0;
	\item Declares a variable \texttt{x} in the Python interpreter and assigns it the value of OpenFOAM variable \texttt{a};
	\item Calls the Python function \texttt{double\_value} with the Python variable \texttt{x} as an argument, and assigns the result to the Python variable \texttt{y};
	\item Copies the value of the Python \texttt{y} variable to a new OpenFOAM variable \texttt{b}.
\end{itemize}

Table {\ref{Evolution_variables_each_line}} shows the variables that in the OpenFOAM/C++ and Python scopes, and how their values change as each line of code is executed in sequence.
\begin{table}[htb]
\centering
\begin{tabular}{lll}
\toprule
\textbf{Executed OpenFOAM/C++ code}                                         & \textbf{OpenFOAM/C++ variables} & \textbf{Python variables} \\ 
\midrule
const scalar a = 2.0; & a = 2.0 & \\
scope{[}``x"{]} = a; & a = 2.0 & x = 2.0 \\
py::exec(``y = double\_value(x)", scope) & a = 2.0 & x = 2.0, y = 4.0  \\
const scalar b = scope{[}``y"{]};                & a = 2.0, b = 4.0 & x = 2.0, y = 4.0 \\
\bottomrule
\end{tabular}
\caption{Evolution of the variables values as each line in the C++ code is executed.} \label{Evolution_variables_each_line}
\end{table}

%--------------------------------------------------------------------------------------%
%--------------------------------------------------------------------------------------%
\section{Test Cases} {\label{829348}}
%--------------------------------------------------------------------------------------%
%--------------------------------------------------------------------------------------%

This section presents three complementary test cases that demonstrate the methodology proposed in Section \ref{sec:approach}:

\begin{enumerate}
	\item \textbf{Python velocity profile boundary condition}: An OpenFOAM wrapper boundary condition is presented, which uses a Python script - supplied at run-time - to define a temporally-spatially varying velocity profile;
	\item \textbf{Python heat transfer prototyping solver}: A wrapper OpenFOAM heat transfer solver is presented, which invokes a run-time Python script to solve the governing equations at each time step; this case shows how new solvers could be more quickly prototyped using a combination of OpenFOAM and Python;
	\item \textbf{Field calculations using Python} A wrapper solids4foam \cite{cardiff2018open}  mechanical constitutive law is presented to demonstrate how a \texttt{volSymmTensorField} stress field can be calculated using Python as a function of a \texttt{volSymmTensorField} strain field. Analytical expressions are compared with TensorFlow Keras \cite{chollet2015keras} neural networks, with an emphasis placed on efficiency. It is expected that the approach could be easily adapted to any other field calculation.
\end{enumerate}
These cases and the code to run them are publicly available at \url{https://bitbucket.org/ScimonUCD/pybind11foam/src/master/}.

%--------------------------------------------------------------------------------------%
\subsection{Python velocity profile boundary condition} {\label{184778}}
%--------------------------------------------------------------------------------------%
This case demonstrates how the general methodology presented in Section \ref{sec:approach} can be used to create a Python-based boundary condition in OpenFOAM.
In this case, a velocity profile boundary condition is created, where a Python script calculates the patch velocities as a function of spatial coordinates and time.
The boundary condition, named \texttt{pythonVelocity}, is a Dirichlet condition and hence derives from the \texttt{fixedValueFvPatchField} boundary condition class.
The central concepts of the \texttt{pythonVelocity} boundary condition - related to the private data and \texttt{updateCoeffs} function - are are described here.

%--------------------------------------------------------------------------------------%
\textbf{Boundary condition private data}
%--------------------------------------------------------------------------------------%
Two private data objects are stored in the boundary condition class:
\begin{itemize}
        \item The name of the Python script (\texttt{fileName pythonScript\_}) which is read at run-time;
        \item The Python interpreter scope object (\texttt{py::object scope\_}) used for transferring data from OpenFOAM to Python, evaluating Python functions, and transferring data back from Python to OpenFOAM.
\end{itemize}

%--------------------------------------------------------------------------------------%
\textbf{Boundary condition \texttt{updateCoeffs} function}
%--------------------------------------------------------------------------------------%
%This was achieved by modifying the \texttt{updateCoeffs()} function of the boundary condition.
%As explained next:
For each finite volume boundary condition in OpenFOAM (those derived from \texttt{fvPatchField}), the \texttt{updateCoeffs} function is called as a preliminary step to the linear solver call.
Consequently, the \texttt{updateCoeffs} function is an appropriate location to update patch velocities as a function of spatial position and time.
As per Section \ref{sec:approach}, there are four essential operations that need to be included in the custom boundary condition:
\begin{itemize}
        \item Within the boundary condition constructor, initialise the Python interpreter and load the Python file;
        \item Within \texttt{updateCoeffs}, transfer the list of face-centre coordinates and time value to Python;
        \item Within \texttt{updateCoeffs}, call the Python script to calculate the list of new face velocities;
        \item Within \texttt{updateCoeffs}, transfer the list of new face velocities back to OpenFOAM.
\end{itemize}

In the boundary condition constructor, the Python interpreter is initialised and the Python script is read:
\begin{lstlisting}[language=C++]
// Initialise the Python interpreter
py::initialize_interpreter();
scope_ = py::module_::import("__main__").attr("__dict__");

// Evaluate the Python script to import modules
py::eval_file(pythonScript_, scope_);
\end{lstlisting} 

Next, in the \texttt{updateCoeffs} function, the patch face-centre coordinate vectors and current time are transferred to the Python scope:
% First, a reference to the face centres is taken:
\begin{lstlisting}[language=C++]
// Convert the face-centre position vectors to a std::vector
const vectorField& C = patch().Cf();
std::vector<std::vector<scalar>> inputC(C.size());
forAll(C, faceI)
{
    inputC[faceI] = std::vector<scalar>(3);
    for (int compI = 0; compI < 3; compI++)
    {
        inputC[faceI][compI] = C[faceI][compI];
    }
}

// Convert std vector to a C++ NumPy array
const py::array inputPy = py::cast(inputC);

// Transfer the C++ NumPy array to a NumPy array in the Python scope
scope_["face_centres"] = inputPy;
scope_["time"] = db().time().value();
\end{lstlisting}

The run-time Python script is then called to calculate the new velocities as a function of the face-centre coordinate vectors (\texttt{face\_centres}) and the current time (\texttt{time}):
\begin{lstlisting}[language=C++]
py::exec("velocities = calculate(face_centres, time)\n", scope_);
\end{lstlisting}
The output \texttt{velocities} list exists in the Python scope.
Note that the only requirement of the run-time loadable Python script is that it contains the definition of a function called \texttt{calculate} which accepts two arguments, one a NumPy array of vectors and the other a double, and returns a NumPy array of vectors.
A suitable example Python function, included in the accompanying \texttt{setInletVelocity.py} file, is:

\begin{lstlisting}[language=Python]
import numpy as np

def calculate(face_centres, time):
    # Initialise result
    result = np.zeros(shape = face_centres.shape)
    # Calculate values using the x coordinates and time
    x = face_centres[:, 0]
    result[:, 0] = np.sin(np.pi*time)*np.sin(40*np.pi*x)
    
    return result
\end{lstlisting} 
where the expression in this case varies the X component of velocity as a function of the X component of the face-centre coordinate vector ($x$) and time ($t$), according to:
\begin{align} %\label{eqn:sinProfile}
u(x, t) = \sin(\pi t)\sin(40 \pi x) \label{eqn:sinProfile}
\end{align}

The calculated velocities are then retrieved from the Python scope and converted into an OpenFOAM field:
\begin{lstlisting}[language=C++]
// Convert the Python velocities to a C++ NumPy array
const py::array outputPy = scope_["velocities"];

// Convert the C++ NumPy array to a C++ std vector
const std::vector<std::vector<scalar>> outputC =
    outputPy.cast<std::vector<std::vector<scalar>>>();

// Convert the C++ std vector to an OpenFOAM field
vectorField velocities(patch().size());
forAll(velocities, faceI)
{
    for (int compI = 0; compI < 3; compI++)
    {
        velocities[faceI][compI] = outputC[faceI][compI];
    }
}
\end{lstlisting} 

Finally, the velocity values are set on the patch:
\begin{lstlisting}[language=C++]
fvPatchField<vector>::operator==(velocities);
fixedValueFvPatchVectorField::updateCoeffs();
\end{lstlisting} 

%--------------------------------------------------------------------------------------%
\textbf{Verification of the proposed boundary condition}
%--------------------------------------------------------------------------------------%
The proposed \texttt{pythonVelocity} boundary condition is verified on the classic \texttt{cavity} tutorial case \cite{lid_driven_cavity_tutorial}, where it is used to set the velocity on the upper moving wall patch.
The only modifications made to the tutorial were to change the velocity boundary condition for the moving wall patch in the initial time (in \texttt{0/U}):
\begin{lstlisting}[language=C++]
movingWall
{
    type            pythonVelocity;
    pythonScript    "$FOAM_CASE/setInletVelocity.py";
    value           uniform (0 0 0);
}
\end{lstlisting}
where the \texttt{setInletVelocity.py} script is placed in the case directory.

\begin{figure}[htb]
	\begin{center}
	\includegraphics[width=0.90\columnwidth]{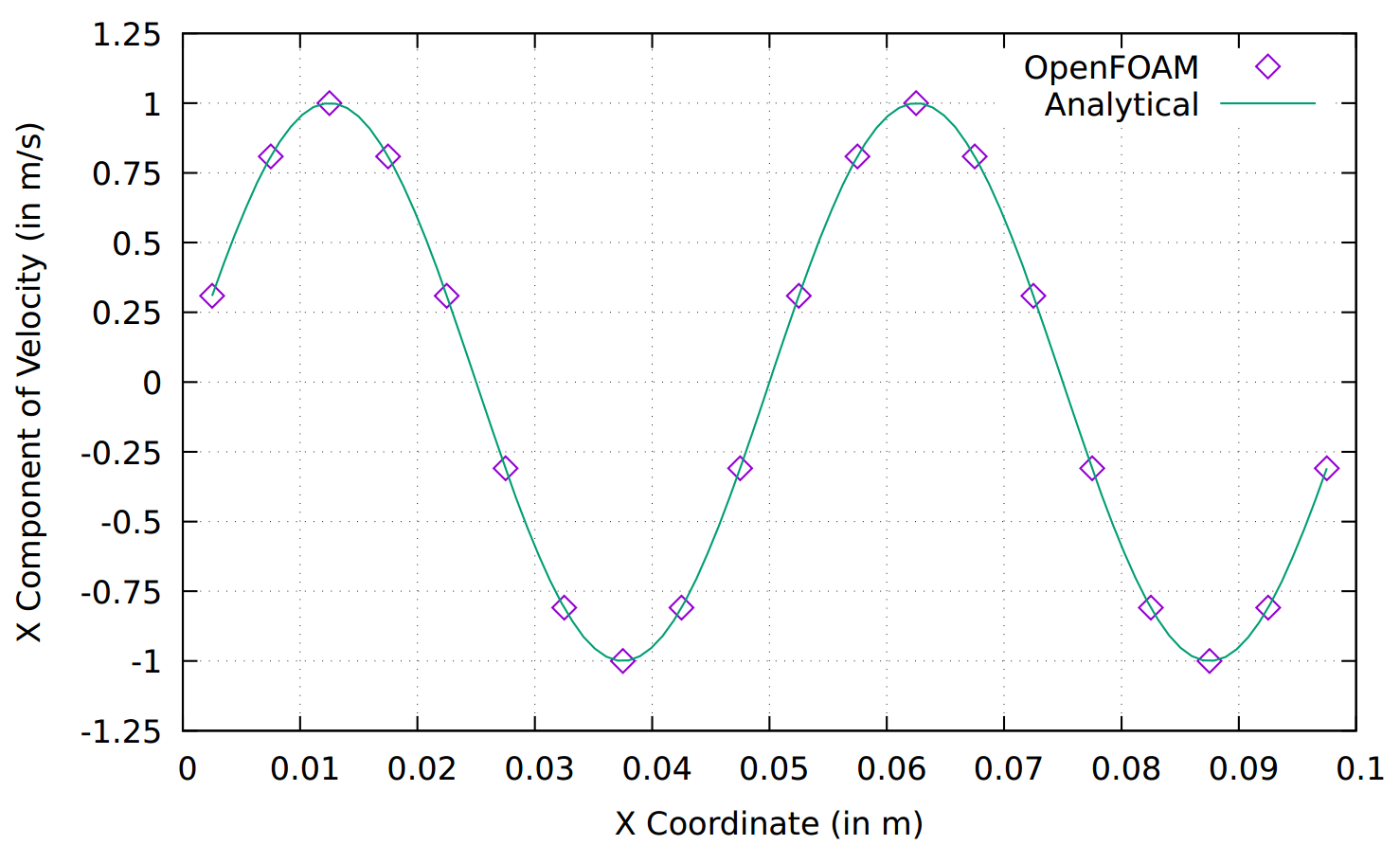}
	\caption{{Velocity along the top boundary at $0.5$ s {\label{508357}}}}
\end{center}
\end{figure}

\begin{figure}[htb]
	\begin{center}
	\includegraphics[width=0.90\columnwidth]{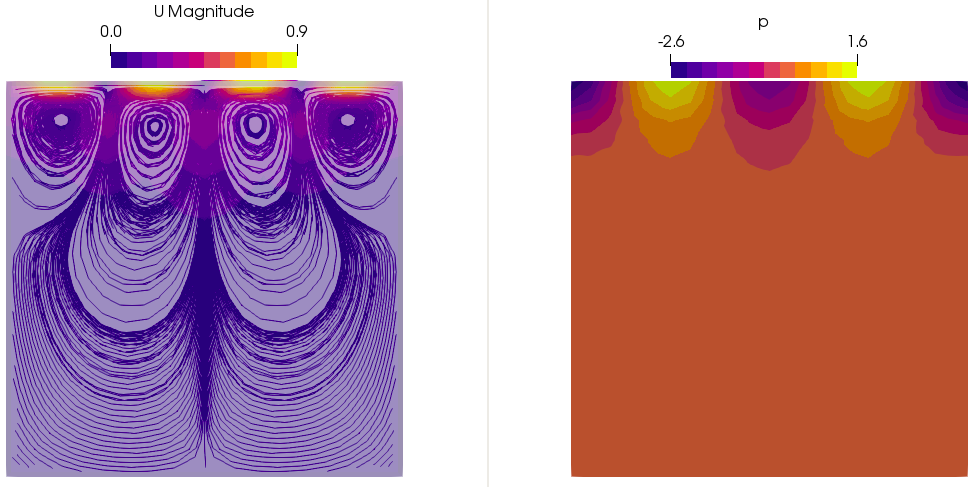}
	\caption{{Velocity streamlines in m s$^{-1}$ (left) and pressure field in Pa m$^3$ kg$^{-1}$ (right) for the cavity case, where the velocity on the upper wall is prescribed via a Python script.
	{\label{701040}} 
	}}
\end{center}
\end{figure}

Running the case produces the velocity profile on the \texttt{movingWall} patch at 0.5 s as shown in Figure {\ref{508357}}, where the expected analytical expression is given for comparison.
The corresponding velocity streamlines and pressure distribution are shown in Figure {\ref{701040}}.

%--------------------------------------------------------------------------------------%
\subsection{Python heat transfer prototyping solver} {\label{304885}}
%--------------------------------------------------------------------------------------%
This case demonstrates one of Python's main advantages over C++: fast prototyping.
To do this, as an example, the classic \texttt{laplacianFoam} solver is modified such that a run-time selectable Python script is passed mesh, material and time information and is expected to calculate the temperature field.
In this case, a simple explicit finite difference method is implemented in Python, however, the approach could be modified to accommodate the prototyping any particular solution strategy or component of it.

As in the first test case, there are three main components added to the OpenFOAM code: data transfer from OpenFOAM to Python; execution of the Python code; transfer of the result data back from Python to OpenFOAM.

%--------------------------------------------------------------------------------------%
\textbf{Data transfer from OpenFOAM to Python}
%--------------------------------------------------------------------------------------%
Following the approach in Section \ref{sec:approach}, OpenFOAM data is mapped, where relevant, to a corresponding Python NumpPy array data.
In this case, two pieces of data are passed to the Python function:
\begin{itemize}
	\item A parameter \texttt{gamma}, which encompasses diffusivity, mesh spacing and pseudo-time-step data:
\begin{lstlisting}[language=C++]
scope["gamma"] =
scalar
(
    DT*runTime.deltaT()*Foam::pow(max(mesh.deltaCoeffs()), 2)
).value();
\end{lstlisting}
		
	\item The temperature field, including boundary condition values, which are assumed here to be \texttt{fixedValue}-type (Dirichlet) conditions:
\begin{lstlisting}[language=C++]
// Set boundary cell values to be equal to the boundary values as we
// will use a finite difference method in the Python script
// These boundary cells will be the boundary nodes in the finite
// difference code
scalarField& TI = T.primitiveFieldRef();
forAll(T.boundaryFieldRef(), patchI)
{
    const labelUList& faceCells =
        mesh.boundary()[patchI].faceCells();
    forAll(faceCells, faceI)
    {
        const label cellID = faceCells[faceI];
        TI[cellID] = T.boundaryFieldRef()[patchI][faceI];
    }
}

// Convert T field to C++ std vector
std::vector<scalar> inputC(TI.size());
forAll(TI, cellI)
{
    inputC[cellI] = TI[cellI];
}

// Convert std vector to NumPy array
const py::array inputPy = py::cast(inputC);

// Assign inputs to Python scope
scope["T"] = inputPy;
\end{lstlisting}

\end{itemize}

%--------------------------------------------------------------------------------------%
\textbf{Perform Python calculations}
%--------------------------------------------------------------------------------------%
As in the first test case, the Python script is selected at run-time, however, in this case it is expected that the provided Python script contains a \texttt{calculate} function, which takes two arguments - the temperature field \texttt{T} with boundary conditions, and the \texttt{gamma} variable - and returns the new temperature field.
The function is called in the OpenFOAM solver as:
\begin{lstlisting}[language=C++]
py::exec("T = calculate(T, gamma)\n", scope);
\end{lstlisting} 

The Python function in this case employs a simple steady-state explicit finite difference approach to solve the Laplace equation:
\begin{lstlisting}[language=Python]
def calculate(T, gamma):

    # Get number of cells in x and y directions
    N = T.shape[0]
    Nx = np.sqrt(N).astype(int)
    Ny = Nx

    # Use explicit finite difference method to update the non-boundary cells
    for i in range(1, Nx - 1):
        for j in range(1, Ny - 1):
            T[i*Ny + j] = \
                gamma*(T[i*Ny + j + 1] + T[i*Ny + j - 1] \
                    + T[(i + 1)*Ny + j] + T[(i - 1)*Ny + j] \
                    - 4*T[i*Ny + j]) + T[i*Ny + j]

    return T
\end{lstlisting}

%--------------------------------------------------------------------------------------%
\textbf{Data transfer from Python to OpenFOAM}
%--------------------------------------------------------------------------------------%
The temperature field results are then extracted from the Python scope back to the OpenFOAM solver as: 

%\hl{also we should rename the OF solver as it does not have to use the FD method, as it is decided by Python}
\begin{lstlisting}[language=C++]
// Transfer Python NumPy array to C++ NumPy array
const py::array outputPy = scope["T"];

// Convert C++ NumPy array to C++ std vector
const std::vector<scalar> outputC =
    outputPy.cast<std::vector<scalar>>();

// Convert C++ std vector to OpenFOAM field
forAll(TI, cellI)
{
    TI[cellI] = outputC[cellI];
}
\end{lstlisting}

%--------------------------------------------------------------------------------------%
\textbf{Results}
%--------------------------------------------------------------------------------------%
The proposed prototyping solver is demonstrated on a steady-state heat conduction problem with a 2-D square geometry ($0.1 \times 0.1$ m) and $20 \times 20$ cells - the geometry and mesh from the \texttt{cavity} tutorial.
The diffusivity is $4 \times 10^{-5}$ m$^2$ s$^{-1}$.
The left, bottom, and right boundaries are set to 273 K, while the top boundary is set to 373 K.
The Python script is specified in \texttt{constant/transportProperties} as:
\begin{lstlisting}[language=C++]
pythonScript    "$FOAM_CASE/calculateT.py";
\end{lstlisting}
where \texttt{calculateT.py} is the Python explicit finite difference code shown above.

\begin{figure}[htb]
	\begin{center}
	\includegraphics[width=0.95\columnwidth]{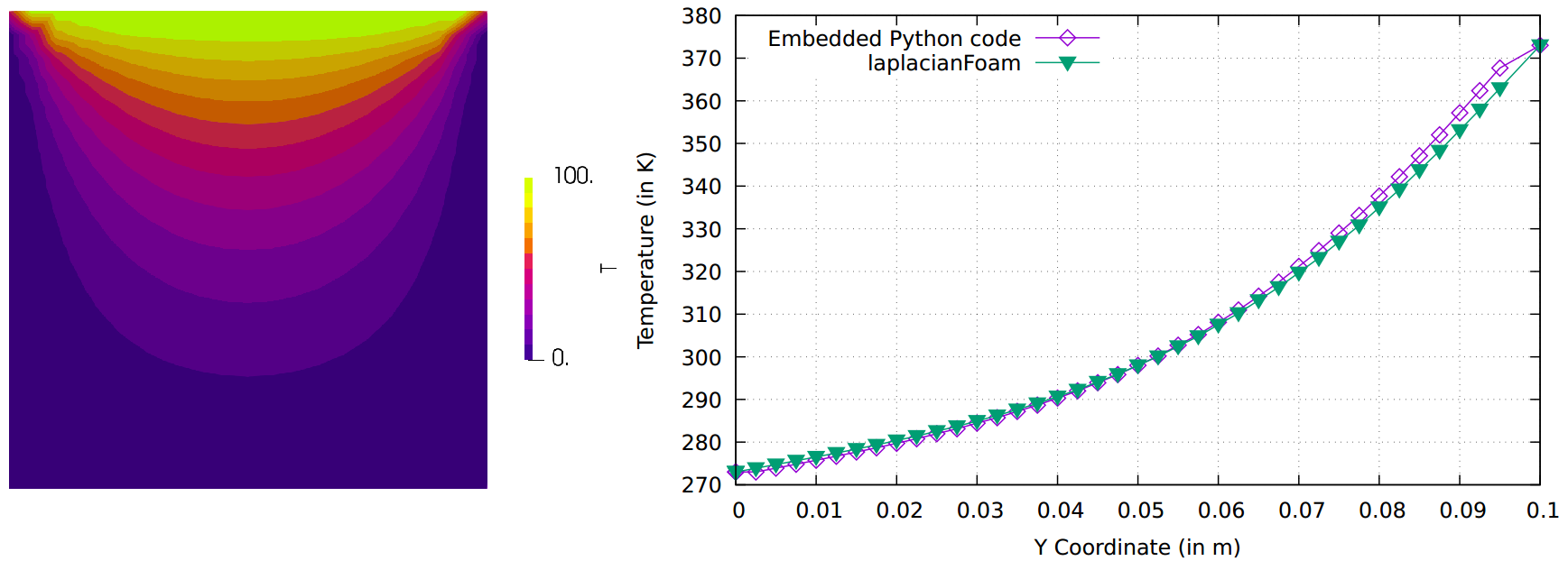}
	\caption{{Steady-state temperature distribution as calculated by a Python finite difference code embedded in an OpenFOAM solver (left) and temperature profile along a vertical line from the centre of the bottom patch to the centre of the top patch, compared with the predictions from \texttt{laplacianFoam} (right). {\label{533014}} }}
	\end{center}
\end{figure}

The final converged temperature field is shown in Figure \ref{533014}(a). For comparison, the temperature along a vertical line from the centre of the bottom patch to the centre of the top patch are compared with the results as calculated by \texttt{laplacianFoam} (Figure \ref{533014}(b)). As expected the predictions are similar, with the differences primarily attributed to the Python finite difference code enforcing boundary conditions in the cell-centres of cells adjacent to the boundaries.

%--------------------------------------------------------------------------------------%
\subsection{Field calculations using Python} 
\begin{figure}[htb]
	\begin{center}
	\includegraphics[width=0.5\textwidth]{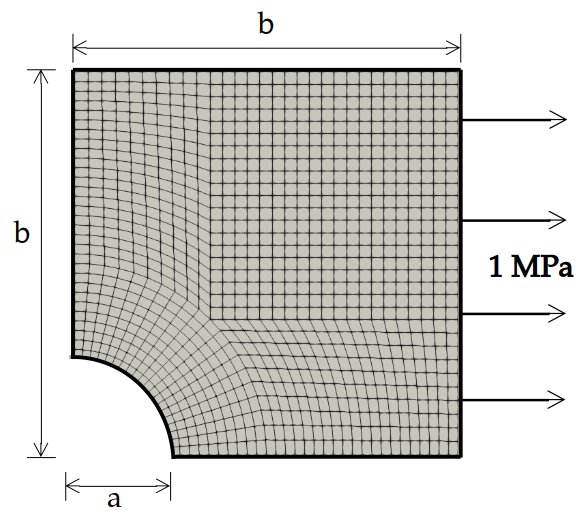}
	\caption{Geometry of the spatial computational domain for the flat plate with a circular hole (a = 0.5 m, b = 2 m, $E = 200$ GPa, $\nu$ = 0.3, $t_x$ = 1MPa). {\label{fig:plateHole}} }
	\end{center}
\end{figure}

{\label{sec:test_case_field_calculations}}
%--------------------------------------------------------------------------------------%
The purpose of this final test case is to demonstrate how to perform field calculations using the embedded Python interpreter;
specifically, this case shows how this can be done using machine learning models implemented in TensorFlow/Keras, scikit-learn, and Python in general.
A solid mechanics problem is chosen, where the stress tensor in each cell is calculated as a function of the displacement gradient via a run-time selectable constitutive mechanical law.
This approach could be readily applied to similar field calculations, such as the calculation of turbulent quantities, thermo-physical property fields, or even for discrete differential operators.
The case (Figure \ref{fig:plateHole}) consists of a plate, with a circular hole in the centre, loaded by uniform tension \(t_x\ =\) 1 MPa on the right boundary, while symmetry conditions are applied to the left and bottom boundaries.
Two quadrilateral meshes were considered: a coarse mesh with 1 000 cells (Figure \ref{fig:plateHole}), and a finer mesh with 400 000 cells.
 All the simulations were solved for 1 time step, small strains were assumed, and the inertia and gravity terms were neglected.
 A Hookean linear elastic material is assumed with a Young's modulus of $E = 200$ GPa and a Poisson's ratio of $\nu = 0.3$.

At each outer iteration within a time-step, the stress tensor $\boldsymbol{\sigma}$ in each cell is calculated as a function of the strain tensor $\boldsymbol{\epsilon}$, according to Hooke's law as
\begin{align} \label{eqn:HookeanLaw}
\boldsymbol{\sigma}	= 2\mu \boldsymbol{\epsilon} + \lambda \, \text{tr} \left(  \boldsymbol{\epsilon} \right) \textbf{I}
\end{align}

where the strain tensor $\boldsymbol{\epsilon}$ is the symmetric component of the displacement gradient $\boldsymbol{\nabla} \boldsymbol{d}$:
\begin{align}
\boldsymbol{\epsilon} = \frac{1}{2} \left[ \boldsymbol{\nabla} \boldsymbol{d} + \left(\boldsymbol{\nabla} \boldsymbol{d}\right)^T \right]
\end{align}

For demonstration purposes, the calculation of the stress tensor in each cell (Equation \ref{eqn:HookeanLaw}) is performed using seven alternative approaches:
\begin{enumerate}
%\tightlist
	\item \textbf{\texttt{solids4foam} reference case}: this simulation uses the native \texttt{solids4foam}/OpenFOAM implementation and does \emph{not} use Python.
	\item \textbf{Analytical expression in Python} (Analytical-Python): the strain tensor is passed to the Python interpreter; the stress is then calculated in Python as per Equation \ref{eqn:HookeanLaw} and returned to OpenFOAM; in this case, no machine learning methods are used.
	\item \textbf{Neural network base case} (NN-Base): as per approach 2, the strain is passed to Python, but in this case the stress is called using an artificial neural network created in Keras, which has been trained on data from Equation \ref{eqn:HookeanLaw}. The prediction method uses the built-in \texttt{.predict} Keras function. \\
	\emph{Approaches 4 to 7 use the same trained neural network as in approach 3, however, alternative neural network prediction methods are examined.}
	\item  \textbf{Neural network functional model} (NN-Functional): the neural network is constructed using the Keras Functional API, as opposed to the Sequential model used in approach 3 \cite{chollet2015keras}.
	\item \textbf{Neural network with eager mode disabled} (NN-Eager-Disabled): the so-called \emph{eager execution} mode \cite{noauthor_NoEager_nodate} of TensorFlow -- which Keras is built-on -- is disabled.
	\item \textbf{Neural network using NumPy} (NN-NumPy): the Keras neural network is fully translated to pure NumPy array calculations, breaking the dependency with Keras and TensorFlow.
	\item \textbf{Neural network using NumPy-Numba} \cite{lam2015numba}  (NN-NumPy-Numba): Numba is an open source just-in-time compiler that translates a subset of Python and NumPy code into fast machine code \cite{noauthor_NumbaHighPerformanceCompiler_nodate}. Here, Numba is applied to approach 6.
\end{enumerate}
Each of these seven approaches are implemented in their own \texttt{python\_file.py} script, loaded at run-time.
Details of these Python implementations are given in Appendix \ref{app:fieldCalc}.

The hole-in-a-plate case is solved twice using each of the seven approaches above:
in the first solution, the strain tensor is passed to the Python interpreter on a cell-by-cell basis, whereas in the second solution, the entire OpenFOAM strain field (all cells) is transferred in one go to the Python interpreter, as discussed in Section \ref{TransferOFPython}.

\begin{figure}[htb]
	\centering
	\subfigure[$\sigma_{xx}$ stress from the top of the hole to the top boundary.]
	{
		\includegraphics[height=0.39\textwidth]{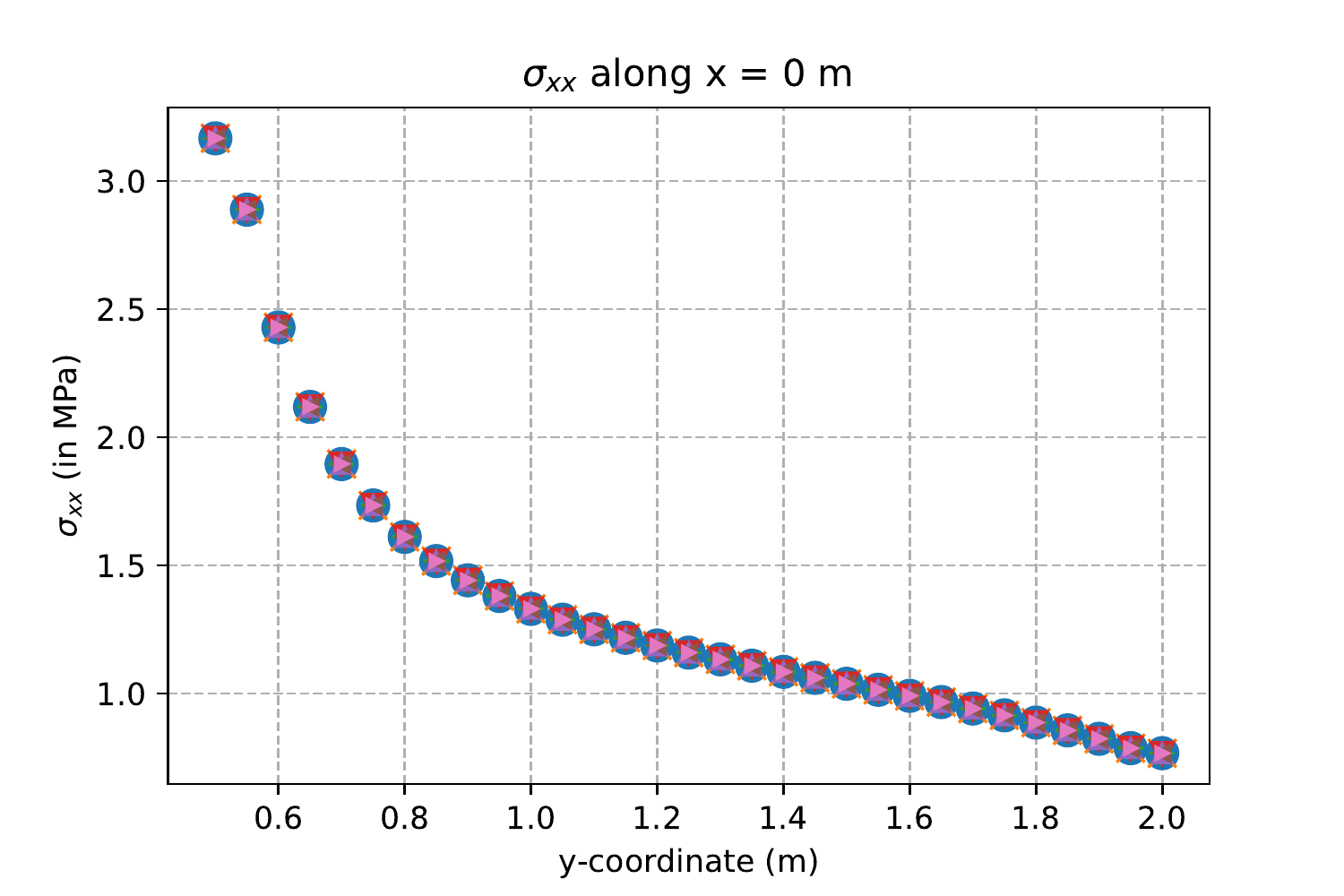}
		\label{fig:resultsPythonLinearElastic}
	}
	\subfigure[Von Mises stress distribution for the solids4foam reference case.]
	{
		\includegraphics[height=0.39\textwidth]{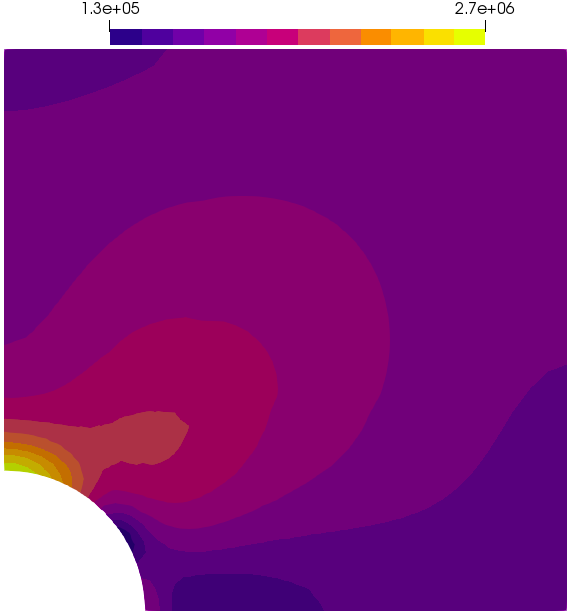}
		\label{fig:vonMisesStressResults}
	}
	\caption{Stress predictions for the hole-in-a-plate case}
\end{figure}

%--------------------------------------------------------------------------------------%
\textbf{Performance comparison} 
%--------------------------------------------------------------------------------------%
%3.3.3. Results
The predicted von Mises stress distribution for the reference \texttt{solids4foam} case is shown in Figure \ref{fig:resultsPythonLinearElastic}, while the predicted $xx$ component of the stress tensor $\boldsymbol{\sigma}_{xx}$ along the vertical line $x = 0$ is shown for all seven approaches in Figure \ref{fig:vonMisesStressResults}.
At this scale, the prediction from all approaches closely agree. Table \ref{MeanErrorMagnitude} lists the $L_\infty$ norm and average $L_2$ norm of the differences with the reference case for the displacement magnitude and von Mises stress fields; 
the maximum errors in the von Mises stress ($<800$ MPa) and displacement magnitude ($<3$e-9 m) are small in comparison to the maximum von Mises stress (2.7 MPa) and maximum displacement magnitudes (13 $\mu$m) in the reference solution.
 In addition, the predictions are the same regardless of whether the data is transferred on a cell-by-cell basis or by the entire field, and as such the data is only presented once for each approach.

% \begin{figure}[htb] 
%     %  \begin{subfigure}
%     %      \centering
%     %      \includegraphics[width=\textwidth]{./figures/CtoPya}
%     %      \caption{Forward conversion. Passing an OpenFOAM field to the Python interpreter implies extracting its data to create a C++ std::vector, creating a py::array from it and transferring it to the Python interpreter}
%     %      \label{fig:ForwardConversion}
%     %  \end{subfigure}
%      %\captionsetup[subfigure]
%      %\begin{subfigure}[b]{0.58\textwidth}
% 	 \begin{subfigure}
% 	    \centering
% 	    \includegraphics[height=0.35\textwidth]{./figures/resultspybindLinearElastic.pdf}
% 	    \caption{$\sigma_{xx}$ stress from the top of the hole to the top boundary patch for the tested Hookean laws.}
% 	    %\label{fig:resultsPythonLinearElastic}
%  	 \end{subfigure}
%     %\hfill
% 	 %\begin{subfigure}[b]{0.37\textwidth}
% 	 \begin{subfigure}
% 	    \centering
% 	    \includegraphics[height=0.45\textwidth]{figures/Resulss4f.png}
% 	    %\caption{Von Mises stress field for the reference case obtained by solids4foam.}
% 	    %\label{fig:vonMisesStressResults}
% 	 \end{subfigure}
% 	 \caption{{Stress predictions for the hole-in-a-plate case.}}
% \end{figure}

\begin{table}[htb]
\begin{tabular}{lllll}
\toprule
& \multicolumn{2}{l}{\textbf{Von Mises Stress}} & \multicolumn{2}{l}{\textbf{Displacement}} \\ 
& \textbf{(in Pa)}            & & \textbf{(in m)}         &        \\
& \textbf{$L_2$}            & \textbf{$L_{\infty}$}          
& \textbf{$L_2$}         & \textbf{$L_{\infty}$}         \\ \midrule
\textbf{solids4foam}       & -                      & -                            & -                   & -                          \\
\textbf{Analytical-Python} & 7.12                   & 775                          & 4.91e-11            & 2.50e-9                    \\
\textbf{NN-Base}           & 7.42                   & 788                          & 5.60e-11            & 2.70e-9                    \\
\textbf{NN-Functional}     & 7.70                   & 790                          & 5.87e-11            & 2.80e-9                    \\
\textbf{NN-Eager-Disabled} & 7.42                   & 788                          & 5.60e-11            & 2.70e-9                    \\
\textbf{NN-NumPy}          & 7.42                   & 781                          & 5.62e-11            & 2.70e-9                    \\
\textbf{NN-NumPy-Numba}    & 7.42                   & 781                          & 5.62e-11            & 2.70e-9                    \\ \bottomrule
\end{tabular}
\caption{L$_{\infty}$ norm and mean L$_2$ norm between the predictions from approaches 2-6 with the reference case, where the maximum stress in the reference case is 2.7 MPa and the maximum displacement magnitude is 13 $\mu$m.}
\label{MeanErrorMagnitude}
\end{table}

As expected the predictions are insensitive to the approach, however, there is a wide variation in the timings between the implementations.
Table \ref{ExecutionTime} shows the execution times for all seven implementations on the coarse (1 000 cells) and fine (400 000 cells) meshes, where the `Ratio' is given for timing relative to the reference case.
\begin{table}[htb]
\begin{tabular}{lllllllll}
\toprule
                           %& \multicolumn{8}{c}{\textbf{Execution time (in s)}} \\
                           & \multicolumn{4}{c}{\textbf{Coarse mesh}}                                                                                                                                                                                                                                                                                                                           & \multicolumn{4}{c}{\textbf{Fine mesh}}                                                                                                                                                                                                                                                                                                                            \\ \hline
                           & \multicolumn{1}{l}{\textbf{\begin{tabular}[c]{@{}c@{}}Cell\\ by\\ Cell\end{tabular}}} & \multicolumn{1}{l}{\textbf{\begin{tabular}[c]{@{}c@{}}Ratio\\ \end{tabular}}} & \multicolumn{1}{l}{\textbf{\begin{tabular}[c]{@{}c@{}}Entire \\ Field\end{tabular}}} & \multicolumn{1}{l}{\textbf{\begin{tabular}[c]{@{}c@{}}Ratio\\ \end{tabular}}} & \multicolumn{1}{l}{\textbf{\begin{tabular}[c]{@{}c@{}}Cell\\ by\\ Cell\end{tabular}}} & \multicolumn{1}{l}{\textbf{\begin{tabular}[c]{@{}c@{}}Ratio\\ \end{tabular}}} & \multicolumn{1}{l}{\textbf{\begin{tabular}[c]{@{}c@{}}Entire\\ Field\end{tabular}}} & \multicolumn{1}{l}{\textbf{\begin{tabular}[c]{@{}c@{}}Ratio\\ \end{tabular}}} \\ \hline
\textbf{solids4foam}       & $<1$*                                                                                 & -                                                                                        & $<1$*                                                                                & -                                                                                        & 490*                                                                                  & -                                                                                        & 490*                                                                                & -                                                                                        \\
\textbf{Analytical-Python} & 8                                                                                     & 37.6                                                                                     & 3                                                                                    & 14.2                                                                                    & 1459                                                                                  & 3.0                                                                                     & 531                                                                                 & 1.1                                                                                     \\
\textbf{NN-Base}           & 2189                                                                                  & 10943.6                                                                                 & 18                                                                                   & 91.9                                                                                    & $> 6$ h                                                                      & $> 45$                                                                       & 576                                                                                 & 1.2                                                                                     \\
\textbf{NN-Functional}     & 117                                                                                   & 586.1                                                                                   & 7                                                                                    & 34.3                                                                                    & $> 6$ h                                                                      & $> 45$                                                                       & 587                                                                                 & 1.2                                                                                     \\
\textbf{NN-Eager-Disabled} & 87                                                                                    & 436.6                                                                                   & 7                                                                                    & 34.2                                                                                    & $> 6$ h                                                                      & $>45$                                                                       & 584                                                                                 & 1.2                                                                                     \\
\textbf{NN-NumPy}          & 20                                                                                    & 97.5                                                                                    & 6                                                                                    & 28.0                                                                                    & 791                                                                                   & 1.6                                                                                     & 781                                                                                 & 1.6                                                                                     \\
\textbf{NN-NumPy-Numba}    & 20                                                                                    & 101.7                                                                                   & 6                                                                                    & 32.0                                                                                    & 3574                                                                                  & 7.3                                                                                     & 837                                                                                 & 1.7                                                                                     \\ \bottomrule
\end{tabular}
\caption{Execution time for all the evaluated cases in s (unless stated otherwise), where timings are rounded to the nearest second. The `Ratio' gives the ratio of the time for the given approach to the time for the reference case. *The \emph{cell-by-cell} and \emph{entire field} designations do not apply to the reference solids4foam calculations.}
\label{ExecutionTime}
\end{table}

The first insight to be gained is that passing the data as entire fields is significantly faster than on a cell-by-cell basis;
for the analytical Python method (approach 2), the entire field approach is almost three times faster than the cell-by-cell approach, for both coarse and fine meshes.
The difference is even greater for the neural network approaches, with the entire field approach being orders of magnitude faster than the cell-by-cell approach.

Among the Python-based implementations, the analytical expression (approach 2) is the most efficient, introducing an overhead as small as 8\% when the data is passed as an entire field.
This matches the expected behaviour, considering that the analytical expression involves less floating point operations than any of the examined neural networks.
Comparing the neural network approaches, the neural network base case, the functional model, and the disabled eager methods (approaches 3-5) all perform similarly on the fine mesh, adding approximately an 18\% overhead relative to the reference \texttt{solids4foam} case.
Similar trends are seen for the coarse mesh, although the overhead is significantly greater ($>$600\%);
from this, we can conclude that there is a fixed overhead associated with loading a neural network (and even just the Python interpreter), but this has a lesser effect as the cell count increases.

Examining the final two approaches, which are the non-Keras-based neural networks (NumPy and NumPy-Numba-based), they show a minor advantage for the coarse mesh but actually slow down the solution for the finer mesh.

%--------------------------------------------------------------------------------------%
\textbf{Passing fields by reference} {\label{sec:passing fields_by_reference}}
%--------------------------------------------------------------------------------------%
As a final attempt to reduce the overhead of using the Python interpreter, we will examine the effect of transferring data by reference (\emph{i.e.} by address) rather than by copying the data, as described in Section \ref{passingByCopy}. 
Table \ref{TimingComparisonByValueByReference} compares the results for the pass-by-copy approach presented in the previous section and the pass-by-reference approach, where results are only shown for the fine mesh using the entire field approach.
The pass-by-reference approach can be seen to produce a significant speedup in the timings.
The analytical Python method (approach 2) exhibits essentially the same execution time as the reference solids4foam case.
Similarly, the base neural network approach only shows an overhead of less than 4\%, while the slowest Keras-based neural network (functional API method) only increases the execution time by 7.3\%.
As in the pass-by-copy approach, the NumPy and NumPy-Numba neural networks methods show the slowest neural network performance.

\begin{table}[htb]
\begin{tabular}{lll}
\toprule
                           & \multicolumn{2}{c}{\textbf{Execution Time (in s)}}  \\
                           & \textbf{Pass by Copy} & \textbf{Pass by Reference} \\ \midrule
\textbf{solids4foam}       & 491*                   & 491*                       \\
\textbf{Analytical-Python} & 531                    & 495                        \\
\textbf{NN-Base}           & 576                    & 509                        \\
\textbf{NN-Functional}     & 587                    & 526                        \\
\textbf{NN-Eager-Disabled} & 584                    & 505                        \\
\textbf{NN-NumPy}          & 781                    & 767                        \\
\textbf{NN-NumPy-Numba}    & 837                    & 878                        \\ \bottomrule
\end{tabular}
\caption{Timing results for the pass-by-reference approaches. *The \emph{cell-by-cell} and \emph{entire field} designations do not apply to the native solids4foam calculations.}
\label{TimingComparisonByValueByReference}
\end{table}

%--------------------------------------------------------------------------------------%
%--------------------------------------------------------------------------------------%
\section{Conclusion} {\label{Conclusion}}
%--------------------------------------------------------------------------------------%
%--------------------------------------------------------------------------------------%
This article introduces a general approach for running Python code in OpenFOAM.
This is achieved with the pybind11 header library, which is used to create an instance of the Python interpreter and for exchanging data between OpenFOAM and Python.
The three examples demonstrate the feasibility and efficiency of the presented approach, where the use of machine learning models is demonstrated in the final case.
When dealing with the calculation of fields, passing data cell-by-cell is compared with passing data via the entire field.
In addition, when passing the entire field, passing-by-copy is compared with pass-by-reference.
The key findings of the article:
\begin{itemize}
	\item It is straight-forward to implement functionality such as boundary conditions, solution algorithms and material models in Python via the pybind11 interpreter approach;
	\item The most computationally inexpensive approach is to pass data between OpenFOAM and Python as entire fields by reference, with an overhead of less than a few percent for the best performing feed-forward neural networks with 266 weights, on a mesh with 400 000 cells;
	\item Passing data between OpenFOAM and Python by copy on a cell-by-cell basis can be prohibitively expensive for larger fields, potentially increasing the run-time by multiple orders of magnitude;
	\item The use of pure NumPy neural networks or Numba-enhanced NumPy neural networks show minimal benefits over the native TensorFlow/Keras Python implementations, and are in fact slower in most cases.
\end{itemize}
Although not been demonstrated in article, the presented approach runs in parallel via the standard OpenFOAM MPI approach without any changes; in that case, each processor has its own local copy of the Python interpreter.
It is hoped that the methods presented in this article will expand the universe of Python-based solutions applicable to OpenFOAM, especially those coming from the rapidly evolving machine learning field.
Future work will explore more complex problems related to elasto-plasticity, requiring machine learning models such as recurrent neural networks.

%--------------------------------------------------------------------------------------%
%--------------------------------------------------------------------------------------%
\section{Acknowledgements}
%--------------------------------------------------------------------------------------%
%--------------------------------------------------------------------------------------%

% \emph{To be added before publication}
% Keep commented during submission, as it would affect the blind review
\noindent Financial support is gratefully acknowledged from the Irish Research Council through the Laureate programme, grant number IRCLA/2017/45.

\noindent Additionally, the authors want to acknowledge project affiliates, Bekaert, through the Bekaert University Technology Centre (UTC) at UCD (www.ucd.ie/bekaert), and I-Form, funded by Science Foundation Ireland (SFI) Grant Number 16/RC/3872, co-funded under European Regional Development Fund and by I-Form industry partners. 

\noindent The authors wish to acknowledge the DJEI/DES/SFI/HEA Irish Centre for High-End Computing (ICHEC) for the provision of computational facilities and support (www.ichec.ie).

%This is an example acknowledgements section and should only be included !after! the review process and before publication.

% The OpenFOAM Journal requires the contributions of all authors to be explicitly stated
% As noted below, please see \href{http://img.mdpi.org/data/contributor-role-instruction.pdf}{CRediT taxonomy} for the term explanation.
% Replace the example authors initials (F.A., S.A) below as appropriate

%Please turn to http://img.mdpi.org/data/contributor-role-instruction.pdf (CRediT taxonomy) for the term explanation.

%\authorcontributions{
%Conceptualisation, J.S.;
%methodology, J.S.;
%software, J.S;
%validation, J.S. and P.M.;
%formal analysis, J.S. and P.M.;
%investigation, J.S.;
%resources, P.M.;
%data curation, P.M.;
%writing---original draft preparation, J.S.;
%writing---review and editing, J.S. and P.M.;
%visualisation, J.S.;
%supervision, P.M.;
%project administration, P.M.;
%funding acquisition, P.M.
%All authors have read and agreed to the published version of the manuscript.
%}

%-----------------------------------------------------------------------
%      Start with appendices here
%-----------------------------------------------------------------------
\newpage
% If required, include appendices here
\appendix

% \section{Example appendix}

%--------------------------------------------------------------------------------------%
%--------------------------------------------------------------------------------------%

\section{Python scripts for the linear Hookean laws implemented in section \ref{sec:test_case_field_calculations}} {\label{app:fieldCalc}}

The run-time loaded \texttt{python\_file.py} for each of the six Python methods (approaches 2 to 7) are given below, along with any additional, necessary code.

%--------------------------------------------------------------------------------------%
\textbf{Analytical expression in Python (Analytical-Python)}
%--------------------------------------------------------------------------------------%
\begin{lstlisting}[language=Python]
def predict(strain_tensor):
    stress = np.zeros([strain_tensor.shape[0], 6])
    stress[:, 0] = 2 * lame_2 * strain_tensor[:, 0] \
        + lame_1 * (strain_tensor[:, 0] \
        + strain_tensor[:, 1] + strain_tensor[:, 2])
    stress[:, 1] = 2 * lame_2 * strain_tensor[:, 1] \
        + lame_1 * (strain_tensor[:, 0] \
        + strain_tensor[:, 1] + strain_tensor[:, 2])
    stress[:, 2] = 2 * lame_2 * strain_tensor[:, 2] \
        + lame_1 * (strain_tensor[:, 0] \
        + strain_tensor[:, 1] + strain_tensor[:, 2])
    stress[:, 3] = 2 * lame_2 * strain_tensor[:, 3]
    stress[:, 4] = 2 * lame_2 * strain_tensor[:, 4]
    stress[:, 5] = 2 * lame_2 * strain_tensor[:, 5]
    return stress
\end{lstlisting} 
where \texttt{lame\_1} and \texttt{lame\_2} are the Lamé's constant and shear modulus, respectively.

\begin{figure}[htb]
\begin{center}
\includegraphics[width=0.98\columnwidth]{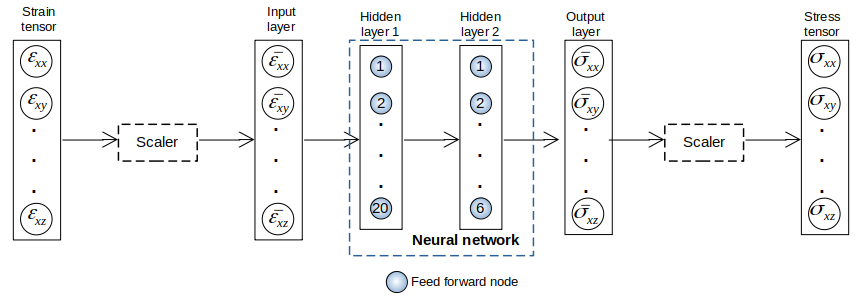}
\caption{{Linear elastic Hookean law based on a neural network. The components of
the strain tensor (\(\epsilon\)) are scaled to the range {[}0,
1{]} (\(\bar{\epsilon}\)) and fed to a neural network. It calculates the
corresponding scaled stress tensor (\(\bar{\sigma}\)) which is scaled
back to its natural range (\(\sigma\)).
{\label{901281}}%
}}
\end{center}
\end{figure}

%--------------------------------------------------------------------------------------%
\textbf{Neural network base case (NN-Base)}
%--------------------------------------------------------------------------------------%
In this case, the mapping between stress and strain from equation ({\ref{eqn:HookeanLaw}}) is provided by a neural network previously trained on Keras.
This neural network has two hidden layers with 20 and 6 feed-forward nodes, respectively.
The activation function of the first hidden layer is ReLu and the other is linear. The entire stress calculation process for a material point is summarised in figure {\ref{901281}}.
At the training phase, both the strain and stress tensors in the training set are scaled into the range {[}0, 1{]} using a linear scaler from the Scikit-Learn \cite{scikit-learn} Python library and fed to the neural network which calculates the corresponding scaled stress tensor which is scaled back to its original range.% with \texttt{y\_scaler.joblib}.\selectlanguage{english}

Therefore, in order to use the resulting neural network, the same scalers need to be available in Python.
The strategy used here was to serialise them once the training phase was completed, place them in the OpenFOAM working directory and load (or deserialise) them in the embedded Python interpreter in OpenFOAM through the \texttt{python\_file.py.}
This is done using the Joblib Python library as shown in the next code:
\begin{lstlisting}[language=Python]
from joblib import load
x_scaler = load('x_scaler.joblib')
y_scaler = load('y_scaler.joblib')
\end{lstlisting} 
where \texttt{x\_scaler.joblib} and \texttt{y\_scaler.joblib} are the strain and stress scalers, respectively.

As in the previous case, the strain-stress mapping is provided by a \texttt{predict} function in the Python code, given by: 
\begin{lstlisting}[language=Python]
def predict(x_new):
    x_new_scaled = x_scaler.transform(x_new)
    x_new_scaled_reshaped = x_new_scaled.reshape(1, x_new.shape[0], 6)
    prediction_scaled = model.predict(x_new_scaled_reshaped)
    prediction_output_scaled = prediction_scaled.reshape(x_new.shape[0], 6)
    prediction = y_scaler.inverse_transform(prediction_output_scaled)
    return prediction
\end{lstlisting} 
In this case, the built-in Keras \texttt{.predict} method is used to perform the
inferences from the neural network.

%--------------------------------------------------------------------------------------%
\textbf{Neural network functional model (NN-Functional)}
%--------------------------------------------------------------------------------------%
When the predictions are performed on a cell-by-cell basis, the \texttt{.predict} method is located within a for loop that iterates over the cells, which is discouraged \cite{predictIsSlower}.
For this reason, the previous case was revisited to modify the \texttt{predict} function.
Instead of using Keras \texttt{.predict}, another neural network is built with the same layers of the previous model using the Functional API from Keras and the inferences are performed by a direct call to the model, as suggested by \cite{predictIsSlower} and shown in the next code: 
\begin{lstlisting}[language=Python]
def predict(x_new):
    x_new_scaled = x_scaler.transform(x_new)
    x_new_scaled_reshaped = x_new_scaled.reshape(1, x_new.shape[0], 6)
    prediction_scaled = model(x_new_scaled_reshaped, training = False)
    prediction_output_scaled = tf.reshape(prediction_scaled, [x_new.shape[0], 6])
    prediction = y_scaler.inverse_transform(prediction_output_scaled)
    return prediction
\end{lstlisting}

%--------------------------------------------------------------------------------------%
\textbf{Neural network with eager mode disabled (NN-Eager-Disabled)}
%--------------------------------------------------------------------------------------%
TensorFlow's eager execution is an imperative programming environment that evaluates operations immediately, without building graphs: operations return concrete values instead of constructing a computational graph to run later \cite{noauthor_NoEager_nodate}.
The following line disables eager execution, as suggested by \cite{predictIsSlower}:
\begin{lstlisting}[language=Python]
tf.compat.v1.disable_eager_execution()
\end{lstlisting} 

Then, another neural network is created and loaded with the parameters used in the previous networks with the code:
\begin{lstlisting}[language=Python]
model = Sequential()
model.add(Dense(units = 20, kernel_initializer = 'he_normal', 
                activation = 'relu', input_shape = (None, 6)))
model.add(Dense(units = 6, kernel_initializer = 'he_normal', 
                activation = 'linear'))
model.compile(optimizer = Adam(lr = 0.01), loss = 'mse')
model.load_weights("DNN.h5")
\end{lstlisting} 
where \texttt{DNN.h5} is the file of the neural network used in the base case.
The prediction step to calculate the stress then procedures the same as in approach 2.

%--------------------------------------------------------------------------------------%
\textbf{Neural network using NumPy (NN-NumPy)}
%--------------------------------------------------------------------------------------%
The same neural network presented in the previous cases was converted to NumPy code using the Python library keras-konverter \cite{keraskonverter}.
First, the parameters of the network are loaded:
\begin{lstlisting}[language=Python]
wb = np.load('DNN_weights.npz', allow_pickle = True)
w, b = wb['wb']
w0, w1 = np.array(w[0], dtype = np.float64), np.array(w[1], dtype = np.float64)
b0, b1 = np.array(b[0], dtype = np.float64) , np.array(b[1], dtype = np.float64)
\end{lstlisting} 

And the predictions from the neural network are given by: 
\begin{lstlisting}[language=Python]
def neural_prediction(x, w0, w1, b0, b1):
    l0 = x.dot(w0) + b0
    l0 = np.maximum(0, l0)
    l1 = l0.dot(w1) + b1  
    return l1
\end{lstlisting} 
which requires the \texttt{predict} function to be changed to:
\begin{lstlisting}[language=Python]
def predict(x):
    x = x_scaler.transform(x)
    x = np.array(x, dtype = np.float64)
    prediction_output_scaled = neural_prediction(x, w0, w1, b0, b1)
    prediction = y_scaler.inverse_transform(prediction_output_scaled)
    return prediction
\end{lstlisting}

%--------------------------------------------------------------------------------------%
\textbf{Neural network using NumPy-Numba (NN-NumPy-Numba)}
%--------------------------------------------------------------------------------------%
Finally, this case aimed to speed execution up by using Numba on the previous NumPy based implementation.
The main difference with the previous case is that Numba is imported and a \texttt{@jit(nopython = True)} decorator is added to the \texttt{neural\_prediction} function:
\begin{lstlisting}[language=Python]
import numba
from numba import jit

@jit(nopython = True)
def neural_prediction(x, w0, w1, b0, b1):
    l0 = x.dot(w0) + b0
    l0 = np.maximum(0, l0)
    l1 = l0.dot(w1) + b1  
    return l1
\end{lstlisting} 
where the prediction step is the same as in the previous approach.

%    Bibliographies can be prepared with BibTeX using IEEEtran style,
\bibliographystyle{IEEEtran}%do not change

\bibliography{Bibliography}

\end{document}